\documentclass[letterpaper,twoside=false,twocolumn,10pt]{scrartcl}

\usepackage{typearea}
\areaset{7in}{9.5in}
\RedeclareSectionCommand[%
    beforeskip=0em,%
    indent=1em%
]{paragraph}

\usepackage{amsmath}
\usepackage{amssymb}
\usepackage{array}
\usepackage{mathtools}
\usepackage{wasysym}
\usepackage{makecell}
\usepackage{pifont}
\usepackage{scalerel}
\usepackage{rotating}
\usepackage{booktabs}
\usepackage{tabularx}
\usepackage[dvipsnames]{xcolor}
\IfFileExists{xurl.sty}{\usepackage{xurl}}{}
\usepackage{enumerate}

\usepackage{float}
\usepackage{xspace}
\usepackage[format=plain]{caption}
\usepackage{subcaption}
\usepackage{multirow}
\usepackage{listings}
\usepackage{tikz}
\usepackage{paralist}
\usepackage[backend=biber,style=ieee,sorting=ynt]{biblatex}
\usepackage{authblk}

\usepackage{placeins}
\usepackage{enumitem}
\setlist[enumerate]{noitemsep}
\setlist[itemize]{noitemsep}

\newif\ifcomments
\commentsfalse

\def\sysname{\textsc{Polytope}\xspace}
\newcommand{\fulltitle}{\sysname: Practical Memory Access Control  \\ for C++ Applications}

\usepackage{hyperref}
\hypersetup{
    pdftitle=\sysname,
    pdfauthor=Anonymized for Submission,
    colorlinks=true,
    linkcolor={red},
    filecolor={red},
    citecolor={blue},
    urlcolor={blue}}
\addtokomafont{title}{\rmfamily}
\addtokomafont{section}{\rmfamily}
\addtokomafont{subsection}{\rmfamily}
\addtokomafont{subsubsection}{\rmfamily}
\addtokomafont{paragraph}{\rmfamily}

\begin{document}

\title{\fulltitle}
\author[$\dag$]{Ioannis Agadakos}
\author[$\ddag$]{Manuel Egele}
\author[$\dag$]{William Robertson}
\affil[$\dag$]{Northeastern University}
\affil[$\ddag$]{Boston University}
\date{}
\maketitle

\begin{abstract}

\textbf{Abstract.}
Designing and implementing secure software is inarguably more important than ever.  However, despite years of research into privilege separating programs, 
it remains difficult to actually do so and such efforts can take years of labor-intensive engineering to reach fruition. 
At the same time, new intra-process isolation primitives make strong data isolation and privilege separation more attractive from a performance perspective. 
Yet, substituting intra-process security boundaries for time-tested process boundaries opens the door to subtle but devastating privilege leaks.
In this work, we present \sysname, a language extension to C++ that aims to make efficient privilege separation accessible to a wider audience of developers. 
\sysname defines a policy language encoded as C++11 attributes that separate code and data into distinct program partitions. 
A modified Clang front-end embeds source-level policy as metadata nodes in the LLVM IR. 
An LLVM pass interprets embedded policy and instruments an IR with code to enforce the source-level policy using Intel MPK.
 A run-time support library manages partitions, protection keys, dynamic memory operations, and indirect call target privileges. 
 An evaluation demonstrates that \sysname provides equivalent protection to prior systems with a low annotation burden and comparable performance overhead. 
 \sysname also renders privilege leaks that contradict intended policy impossible to express.

\end{abstract}


\section{Introduction}
\label{sec:intro}

Software is as complex as it has ever been, and grows increasingly so every year.  Common applications that execute on traditional consumer platforms have SLOC counts in the tens of millions, approaching that of full-featured COTS operating systems,\footnote{Linux~(git revision \texttt{bf05bf16c}) and Mozilla's gecko-dev repository (git revision \texttt{b2c3325b30}) both contain approximately 22.6M SLOC.  Both measurements were produced using tokei.} and can link to anywhere from a handful to hundreds of third-party libraries of diverse origin.  Mobile, web, server, and cloud software all follow a similar trend.  Unfortunately, despite years of software security research, by default all code comprising a program runs with the same privilege on consumer OSes.  Thus, one vulnerability or malicious backdoor in a program results in the full compromise of the entire program, its data, and---usually---all of the resources available to the program from the underlying system.

Over-privilege and ambient authority~\cite{miller_2003_capabilitymythsdemolished} have long been recognized as a secure software design anti-pattern, with privilege separation the traditional mitigation from a design perspective~\cite{brumley_2004_privtransautomaticallypartitioning,openbsdproject_2021_openssh,barth_2008_securityarchitecturechromium,mozillafoundation_2021_electrolysis}.  Privilege separation applies two classic security principles, separation of privilege and least privilege~\cite{saltzer_1975_protectioninformationcomputer}, by first splitting programs into isolated protection domains or \emph{partitions} and then granting each partition the least amount of privilege necessary to execute correctly.  Program partitions are traditionally obtained via process isolation, although recent work has shown how this can be achieved on an intra-process basis using new hardware security extensions such as Intel Memory Protection Keys~(MPK)~\cite{park_2019_libmpksoftwareabstraction,vahldiek-oberwagner_2019_erimsecureefficient,schrammel_2020_donkydomainkeys}. Alternatively, a variety of platform-specific sandboxing frameworks can be used to remove unnecessary privilege from each partition.  For instance, on Linux SECCOMP-BPF~\cite{linuxkerneldevelopers_2020_seccompbpfsecure}, SELinux~\cite{selinuxproject_2020_selinux}, or AppArmor~\cite{bauer_2006_paranoidpenguinintroduction} can all be used in this manner.

However, despite years of academic research into privilege separation~\cite{brumley_2004_privtransautomaticallypartitioning,bittau_2008_wedgesplittingapplications,pearce_2012_addroidprivilegeseparation,mambretti_2016_trellisprivilegeseparation}, designing privilege-separated programs---and especially retrofitting a privilege-separated security model into existing software---remains a labor-intensive and error-prone process.  One might point to notable examples of privilege-separated applications such as OpenSSH~\cite{openbsdproject_2021_openssh}, Google Chromium~\cite{barth_2008_securityarchitecturechromium}, or Mozilla Firefox~\cite{mozillafoundation_2021_electrolysis} as counterpoints to this claim.  However, we argue that these examples, while highly laudable, are at the same time exceptions that prove the rule.  The vast majority of software in existence does not adopt a privilege-separated security architecture, despite its widely-acknowledged benefits.  Anecdotally, this is due to several factors:
\begin{inparaenum}[\itshape (i)\upshape]
    \item design-time overhead inherent in systematically reasoning about how to integrate privilege separation into a software design;
    \item development-time overhead in implementing and testing a privilege-separated design; and,
    \item performance overhead, especially for designs rooted in process isolation that require message passing and context switching for each cross-partition control transfer.
\end{inparaenum}
As one real-world example of these phenomena, it took Mozilla roughly seven years to complete Electrolysis, the internal code name for its effort to migrate Firefox to a process-separated design~\cite{mozillafoundation_2021_electrolysis}.  This data point demonstrates that despite the significant security advantages of privilege separation and strong market pressure to bring Firefox to ``security parity'' with competing browsers, retrofitting privilege separation into an existing program is a massive engineering undertaking. Furthermore, recent work has focused on making privilege separation more efficient by leveraging intra-process hardware memory protection schemes~\cite{park_2019_libmpksoftwareabstraction,vahldiek-oberwagner_2019_erimsecureefficient,schrammel_2020_donkydomainkeys}.  While this work is extremely promising, at the same time we argue that such schemes do not reduce the complexity of adopting privilege separation and can in fact increase the complexity of securing software.  This is due in part to requiring context-sensitive placement of partitioning controls, increasing the cognitive burden on the programmer to verify their correctness~\cite{cant1995conceptual}.

Our vision for this work is to bring the privilege separation security model to a wide range of software by greatly reducing the cost of designing and implementing privilege separation-based security in both legacy and new programs.  A key principle of our approach is to separate the high-level security policy to enforce from the underlying implementation, abstracting away the complexities and potential pitfalls of low-level implementation mechanisms.  A beneficial consequence of this separation is that policies can be enforced using multiple separation backends depending on platform availability or performance requirements.

To this end, we present \sysname, a language extension to C++11 to easily define program partitions, assign code privileges over these partitions, and automatically manage cross-partition control flows.  At development time, \sysname allows developers to express security policies that assign data to partitions and privileges to code.  A compiler pass automatically instruments programs to enforce these policies, while a support library is responsible for managing partition privileges at run-time.  Compile-time control-flow integrity~(CFI) is also applied to ensure that code reuse cannot subvert \sysname isolation policies.  Our prototype implementation of \sysname builds on LLVM and Intel MPK to enable efficient hardware-enforced intra-process privilege separation.  Our evaluation of this prototype shows that a handful of annotations can implement useful security policies for relevant real-world programs while introducing modest overhead in common benchmark applications.
In summary, this paper makes the following contributions:
\begin{itemize}
    \item We show that useful privilege separation policies for C++ programs can be succinctly expressed in source code in terms of program partitions and a combination of implicit and explicit privilege statements on code.
    \item We show that privilege separation policies can be decoupled from low-level enforcement mechanisms, preventing the expression of insecure policies and more generally lifting the burden of ensuring correct usage of these complex mechanisms from application developers.
    \item We present the design and prototype implementation of \sysname, which allows developers to express privilege separation policies in C++ programs.  \sysname automatically implements these policies using compiler-based transformations and a run-time support library.
    \item We demonstrate via empirical evaluation that \sysname-protected programs exhibit comparable overhead to prior intra-process separation approaches, while requiring far less developer effort and significantly reducing the likelihood of developer error.
\end{itemize}

In the interests of promoting open science and reproducibility, we will publish the \sysname prototype source code upon publication of this work.


\section{Motivation and Problem Statement}
\label{sec:motivation}

Many programs can benefit from integrating privilege separation into their security design.  Programs often contain a small set of secrets that they need to protect, such as cryptographic keys.  The majority of a program's code might depend on the (safe) results of computation on those secrets, but on the other hand do not rely on direct access to them.  This follows from long-established software engineering practices that hew to designs embracing modular, loosely-coupled code.

\begin{figure}[h]
  \includegraphics[width=\linewidth]{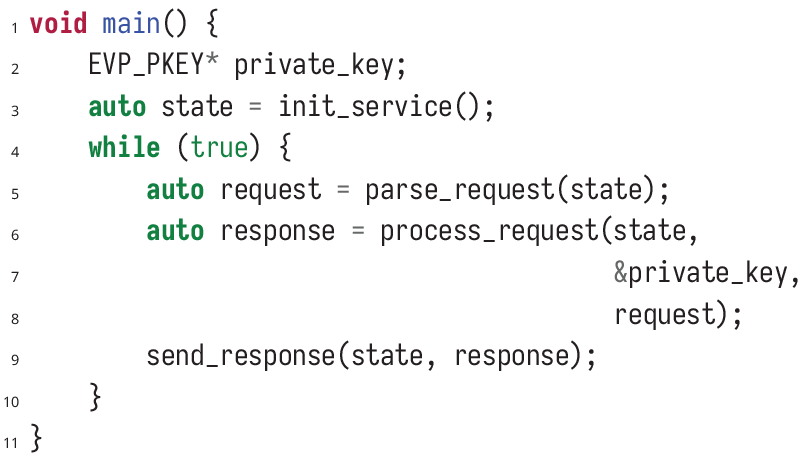}
  \caption{Example program that processes untrusted input and accesses a secret value during execution.  A memory corruption vulnerability anywhere in the program allows an attacker to access all program memory, including \texttt{private\_key}, even though only a relatively small amount of code requires access to that variabl}\label{list:example}
\end{figure}
In this example, the developer is responsible for allocating memory from the correct partition (lines~8-12) and then specifying the code region where protected processing is allowed via matching \texttt{mpk\_begin()} and \texttt{mpk\_end()} invocations.  Similarly to a misplaced unlock operation or memory deallocation, consider the consequences of erroneously placing \texttt{mpk\_end} at line~32.  If \texttt{response.needs\_signature()} evaluates to true, then the intended partition switch out of \texttt{SECRET} would not occur.  This in turn would compromise privilege separation entirely; any subsequent attack would allow an attacker to access \texttt{SECRET} despite the use of libmpk.  This sort of mis-specification constitutes a \emph{privilege leak}.

We note that process-based privilege separation would not be vulnerable to this same error, as \texttt{SECRET} would be isolated in a separate process from the rest of the program.  In this respect, intra-process privilege separation is potentially more brittle than traditional approaches, requiring greater cognitive and engineering overhead to verify that the program implements the intended security policy and does not contain privilege leaks.  We also note that other MPK-based schemes such as ERIM~\cite{vahldiek-oberwagner_2019_erimsecureefficient} and Donky~\cite{schrammel_2020_donkydomainkeys} share this characteristic.

\subsection{Design Goals and Security Properties}
\label{sec:goals}

\begin{figure}[h!]
  \includegraphics[width=\linewidth]{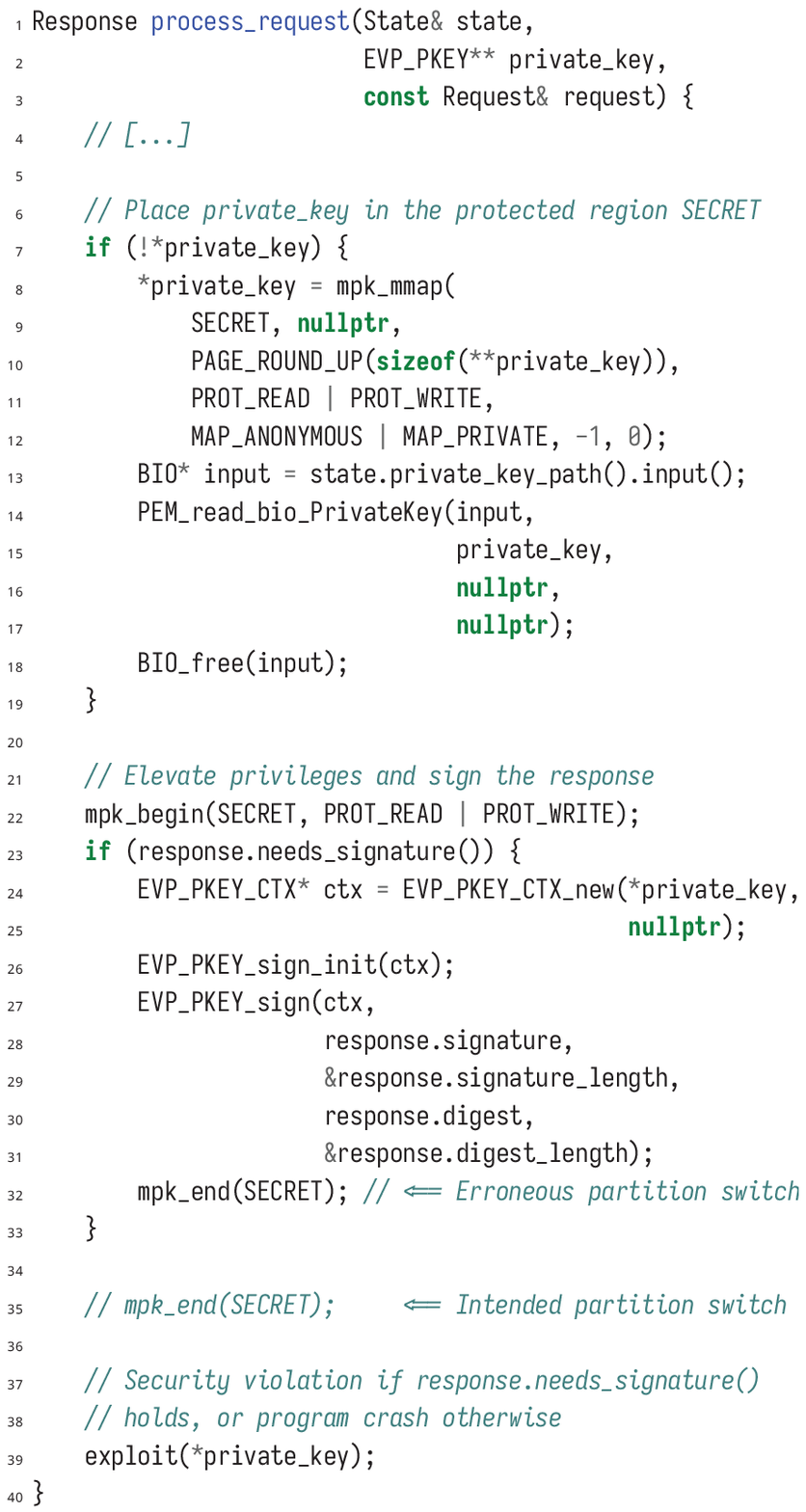}
  \caption{Manual memory allocation and partition switching requires significant development effort and is prone to both security violations and reliability failures.  This usage is loosely based on an example from libmpk~\cite{park_2019_libmpksoftwareabstraction} modulo the introduction of a privilege leak.}\label{list:example_libmpk}
\end{figure}

With the above motivation in mind, our goal with this work is to reduce the cognitive and engineering burden on developers who wish to make use of modern MPK-based privilege separation schemes.  In particular, we aim to achieve the following design goals:

\begin{enumerate}[label=\textbf{(G\arabic*)}]
    \item \label{goal:reasoning} Lift specification of security policy in terms of partition boundaries and privileges from direct expression in code to a level where compilers or other automated tools can reason about and enforce them.
    \item \label{goal:correctness} Prevent the introduction of partition boundary specification errors (cf.~Figure~\ref{list:example_libmpk}).
    \item \label{goal:decoupling} Decouple policy specification from the underlying enforcement mechanism to support alternative separation backends, e.g., process separation or forthcoming, equivalent hardware primitives on other architectures.
    \item \label{goal:performance} Retain the performance characteristics of existing intra-process privilege separation schemes.
\end{enumerate}

\subsection{Threat Model}
\label{sec:threat-model}
We base the threat model that we assume for this work on the standard attacker that possesses an arbitrary read and write primitive in process memory; this is the standard for defenses against memory corruption, such as is assumed for randomization-based or control-flow integrity (CFI) defenses.  However, we consider several classes of attack to be outside our threat model.  We do not defend against side channels.  We also consider hardware vulnerabilities such as row hammering to be out of scope.

We assume that the architecture is capable of enforcing non-writable code pages, and that this capability is correctly applied by the operating system.  We similarly assume that the kernel is trusted and that the hardware correctly implements any necessary intra-process isolation primitives.  We assume that CFI is enforced via compile-time instrumentation, preventing attackers from executing arbitrary sequences of existing code via, e.g., return- or jump-oriented programming (ROP/JOP).  In particular, this prevents attackers from subverting declared \sysname isolation policies.  Finally, we assume that developers do not author code that leaks secrets or privileged capabilities from partitions in violation of the intended security policy.


\section{\sysname}%
\label{sec:design}

\newcommand{\partmain}[0]{\mathsf{main}}
\newcommand{\partlibc}[0]{\mathsf{libc}}
\newcommand{\partlibssl}[0]{\mathsf{libssl}}
\newcommand{\privread}[0]{\mathsf{read}}
\newcommand{\privwrite}[0]{\mathsf{write}}

\begin{figure*}[t]
    \centering
    \includegraphics[width=\linewidth]{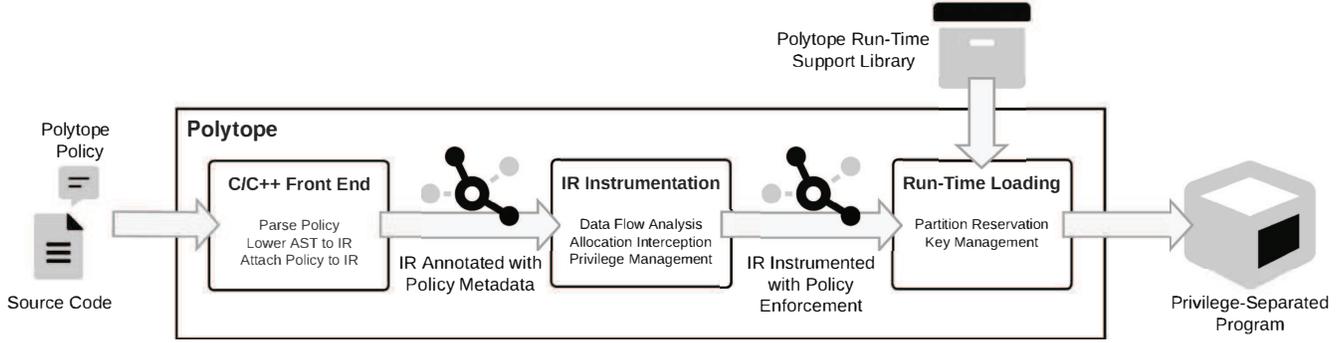}
    \caption{\sysname transforms a traditional monolithic security model for process memory to multiple security policy projections onto process memory.  Source code is enriched with a high-level \sysname policy that defines program partitions, assigns data to each partition, attaches default privileges on partitions to code, and defines allowed control and data flows between each partition.  These policy statements are lowered along with the source code to an intermediate representation (IR) in the form of policy metadata.  A separate pass transforms the annotated IR to enforce the attached policy.  Finally, at execution time a run-time support library is loaded to coordinate reservations of partition memory regions and protection key management required by the separation backend.}
    \label{fig:tesseract_arch}
\end{figure*}

The overarching goal of \sysname is to render privilege-separated security accessible to all C++ programs, whether legacy or newly-authored, by significantly reducing the design and implementation burden associated with manual use of separation primitives.  To accomplish this goal, \sysname moves specification of security policy from pure source code into a declarative access control policy~(\S\ref{sub:abstract_policy}) in the form of coarse policies on translation units along with a small set of explicit policy annotations~(\S\ref{sub:policy_expression}).  A modified Clang front-end lowers policy annotations in source code into the LLVM intermediate representation (IR) as policy metadata~(\S\ref{sub:policy_embedding}).  The policy-annotated IR is then automatically instrumented with the necessary elements to implement the source-level \sysname policy: data partitions and associated memory access privileges, dynamic memory allocations from those partitions, and code privilege management~(\S\ref{sub:policy_instrumentation}).  At execution time, a \sysname run-time support library is loaded alongside the main executable to manage dynamic resources such as hardware memory protection keys~(\S\ref{sub:run_time_support}).  The result is a privilege-separated program that has been hardened against compromise such that
\begin{inparaenum}[\itshape (i)\upshape]
    \item data leakage requires exploitation of a specific associated partition, and
    \item exploiting the remainder of the program does not directly confer access to that data.
\end{inparaenum}

An overview of each stage of the \sysname pipeline is shown in Fig.~\ref{fig:tesseract_arch}.  In the remainder of this section, we present the design of \sysname's security policies, development-time policy expression, instrumentation-based policy enforcement, and run-time support.

\subsection{Security Policies}%
\label{sub:abstract_policy}

At an abstract level, a \sysname policy logically separates a program, which includes both the first-party program and any third-party libraries it uses, into distinct \emph{partitions} that represent security boundaries for program data.  All program data is assigned to a partition, and all code has an associated access control policy that by default allows read+write access to one partition's data and no access to other partitions.  These privileges intersect with any other relevant restrictions on data accesses---for instance, if a variable is marked as immutable (e.g., C++ \texttt{const}) then it will always be read-only regardless of the relevant \sysname privileges.

Several classes of statements comprise \sysname policies: \emph{partition declarations}, \emph{partition assignments}, and \emph{privilege assignments}.  We describe each of these more formally below.

\paragraph{Partition Declarations.}%
\label{par:partition_declarations}
A \sysname security policy first requires declaration of the set of program partitions \(P\).  Each declaration consists simply of a distinct label \(p \in P\).  However, associated default access rights are also defined as the function \(\phi \colon P \mapsto A\), where \(a \in A = \mathcal{P}\left(\left\{\privread,\privwrite\right\}\right)\).  \(\phi\) is total, i.e., it is defined for all partitions.  We also note that \(A\) has the usual partial ordering operator \(<_A\) defined in terms of the set inclusion relation.  For example, \(\emptyset <_A \left\{\privread\right\} <_A \left\{\privread,\privwrite\right\}\), but no ordering is defined for \(\left(\left\{\privread\right\}, \left\{\privwrite\right\}\right)\).

A default access policy is simply the minimum level of access that is granted to any code comprising the program.  Faithful implementation of traditional privilege separation would imply \(\phi\left(p\right) = \emptyset \; \forall p \in P\).  However, other default policies are certainly possible---e.g., specifying default read- or even write-only access instead.

\paragraph{Partition Assignments.}%
\label{par:partition_assignments}

Given a set of declared partitions \(P\) and default access rights \(\phi\), \sysname policies can then assign data as belonging to a particular partition.  That is, a policy defines a data access function \(\alpha \colon V \mapsto P\), where \(V\) is the set of variables defined in the program.  We note that \(\alpha\) is total: \(\alpha\) is defined on all variables, i.e., every variable belongs to a single declared partition.  Assigning each variable to a partition creates a default access policy for the entire program.  This follows from the totality of \(\alpha\) and each partition being endowed with default access rights by \(\phi\).

The definition of \(\alpha\) will naturally depend on the particular program.  However, in principle, the strongest security policy will be obtained by distributing sensitive data across the smallest distinct partitions possible while retaining correct behavior.  However, there might be implementation constraints that limit the total number of partitions available in practice.  A greater number of partitions also implies a greater expected number of cross-partition control flows that carry an associated, implementation-specific cost.

\paragraph{Privilege Assignments.}%
\label{par:privilege_assignments}

With \(P, \phi, \alpha\) in hand, the last component of a \sysname policy is an assignment of privileges to code.  More precisely, a privilege assignment is a function \(\pi \colon S \times P \mapsto A\), where \(S\) is the set of statements comprising the program.  \(\pi\) is also a total function, hence every combination of statement and partition has a concrete privilege defined in terms of \(\mathcal{P}\left(\left\{\privread,\privwrite\right\}\right)\).  An additional constraint on \(\pi\) is that a privilege assignment must always be strictly greater than the default access policy for a partition; that is, \(\phi(p) <_A \pi\left(s, p\right) \;\forall s \in S, p \in P\).

As with \(\alpha\), the definition of \(\pi\) will depend on the particular program.  In this case, traditional privilege separation entails assigning the least privilege sufficient to execute successfully on any well-formed input.  In \sysname's context, this is in part achieved by design by requiring an explicit assignment for code that requires greater access privileges on protected data than the default policy for its parent partition.  This requirement encourages privilege escalation only for the minimal set of statements required for correctness.  However, the onus is currently on the developer to ensure that the least necessary privilege is assigned to this code; automatically proving that minimally sufficient privileges are assigned to the minimal set of statements is a direction left to future work.

A complete \sysname policy for a program is thus the tuple \(\left\langle P, \phi, \alpha, \pi\right\rangle\).

\subsection{Policy Expression}%
\label{sub:policy_expression}

While \sysname policies are conceptually simple in the abstract, a direct translation at development time would be verbose and error-prone.  Thus, \sysname policies are expressed instead as a combination of
\begin{inparaenum}[\itshape (i)\upshape]
    \item \emph{coarse-grained} partition assignment statements on translation units and executable objects such as dynamically-linked shared libraries,
    \item \emph{fine-grained} partition assignment and privilege refinement source code annotations, and
    \item implicit default policies.
\end{inparaenum}
Expressing policies in this manner satisfies our design goal of separating security policy from its implementation~\ref{goal:reasoning}.  Figure~\ref{list:example_tesseract} presents an example policy for the running example program.

\paragraph{Coarse Partition Assignments.}
\label{par:coarse_grained_partition_assignments}
To retain the abstract policy semantics presented in~\S\ref{sub:abstract_policy}, all code and data must belong to a partition.  Since this includes compiled third-party libraries linked by the first-party program that cannot easily be policy-instrumented as described in~\S\ref{sub:policy_instrumentation}, coarse partition assignments take the form of both an (early-binding) translation unit-level source annotation as well as a (late-binding) link-time directive for pre-compiled shared objects.

Translation units can be assigned to a partition using a pragma directive of the form:

\begin{lstlisting}[linewidth=\columnwidth,breaklines=true,language=C++]
    #pragma partition <partition_id> <default_rights>
\end{lstlisting}

\texttt{partition\_id} is simply a non-negative integer label.  This statement has the effect of assigning all global data defined in the translation unit to the named partition.  In addition, all code in the translation unit is assigned \texttt{default\_rights} privileges on the named partition, and no access on other partitions.  All translation units in the first-party program must have an explicitly-assigned partition; however, this can often be achieved with a single pragma directive in a header file.

Link-time partition assignments are semantically similar to translation unit pragmas, but are instead expressed via linker options.  However, as compiled shared objects are not subject to policy instrumentation~(\S\ref{sub:policy_instrumentation}), link-time assignments cannot be refined via fine-grained partition assignments as described below.  If this is desired, then the relevant libraries must be compiled with the \sysname toolchain the same as the first-party program.

\paragraph{Fine-Grained Partition Assignments.}%
\label{par:fine_grained_partition_assignments}
To explicitly assign data to a partition other than that of the enclosing translation unit, \sysname also supports C++ attributes of the following form on variables which may be local, global, or members of a structure or class:

\begin{lstlisting}[linewidth=\columnwidth,breaklines=true,language=C++]
    [[partition(<partition_id>, <default_rights>)]]
\end{lstlisting}

These assignments allow developers to refer to data contained in one partition from another, different partition.

\paragraph{Privilege Refinements.}%
\label{par:annotation_privileges}
Code in one partition can naturally invoke code in another.  By default, cross-partition control flows drop access privileges on the caller's partition and acquire privileges on the callee partition.  If control returns to the caller's partition, then privileges are again exchanged.

However, code can also temporarily \emph{augment} its current privilege set via explicit policy statements rather than implicitly exchanging them via a cross-partition call.  Privilege refinement statements take the form of C++ attributes on various code structures such as statements, blocks, functions, lambdas, and class methods:
\begin{figure}
    \includegraphics[width=\linewidth]{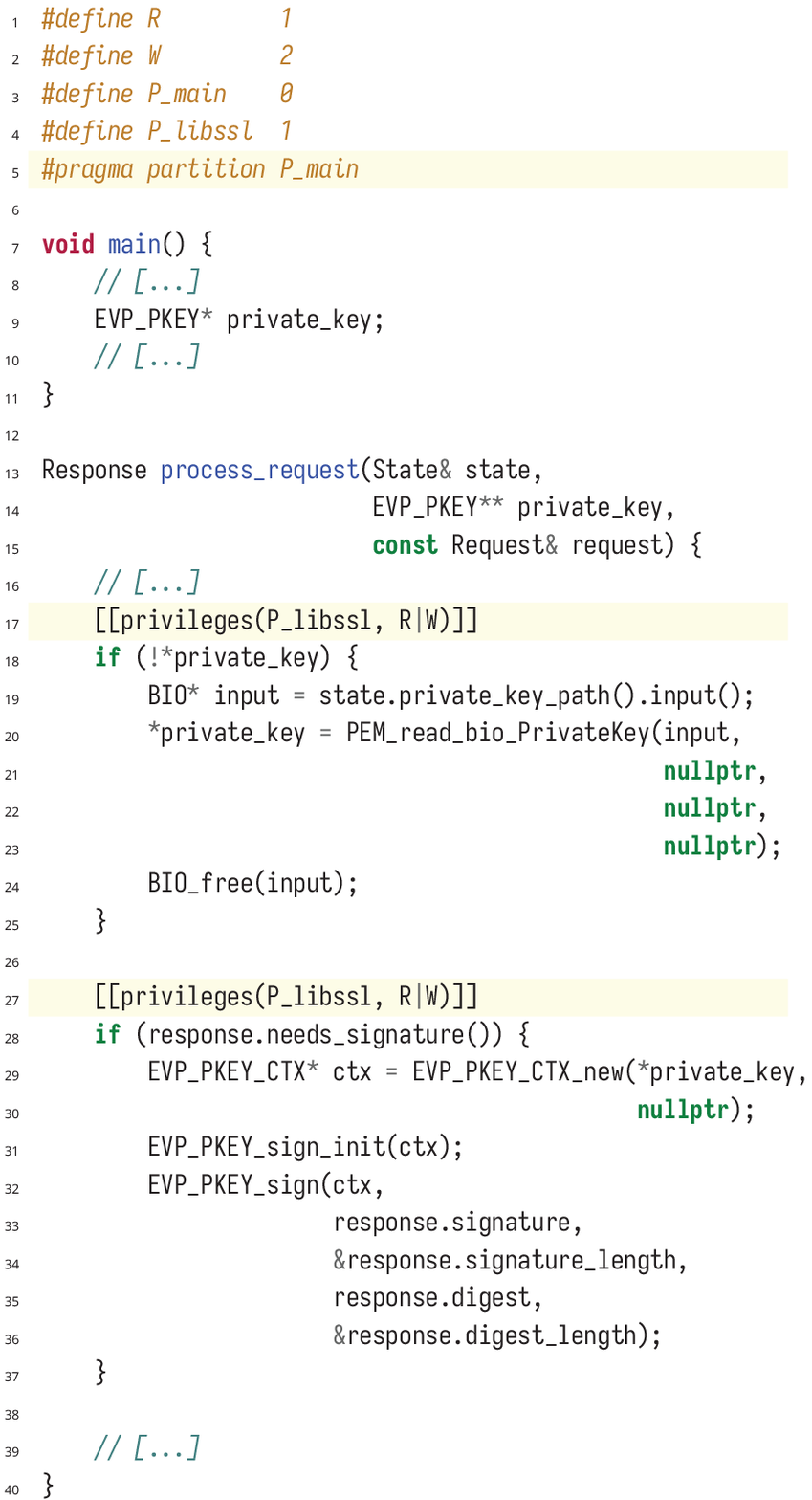}
    \caption{Example program annotated with \sysname policy.  The translation unit is assigned to partition \texttt{P\_main} (l.~5); all global data is in \texttt{P\_main} by default, and all code has read+write privileges on \texttt{P\_main}.  The program is linked against OpenSSL, which is assigned to \texttt{P\_libssl}.  A private key is allocated and initialized within \texttt{P\_libssl} (l.~17-25), and optionally used to sign responses (l.~27-37).  Privileges for the specific code that uses \texttt{private\_key} are raised to include read+write, but outside of that code and \texttt{P\_libssl} the private key is inaccessible.}\label{list:example_tesseract}
\end{figure}

\subsection{Policy Embedding}%
\label{sub:policy_embedding}

\begin{figure}[tp]
    \centering
    \includegraphics[width=\linewidth]{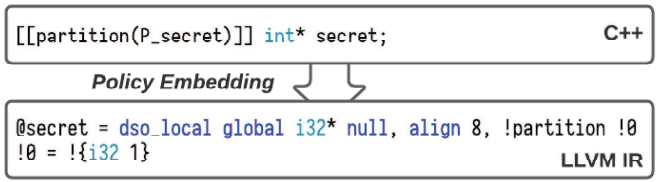}
    \caption{Example of policy embedding.  Here, \texttt{int* secret} is assigned to \texttt{P\_secret}.  \sysname embeds this policy as a metadata node attached to the corresponding LLVM IR variable declaration, where \texttt{P\_secret} has a partition label of 1.}
    \label{fig:embedding}
\end{figure}

The first stage in the \sysname pipeline is responsible for lowering source-level policy into the LLVM IR.  This process occurs as part of the usual Clang front-end lowering of a parsed abstract syntax tree (AST) into LLVM IR.  The embedded policy takes the form of special metadata that is attached to variable declarations and relevant instructions in the IR.  An example of a policy embedding is shown in Figure~\ref{fig:embedding}.

Separating IR policy embeddings from policy enforcement in the following instrumentation stage is a conscious design choice~\ref{goal:decoupling}.  This decoupling allows for multiple enforcement implementations to be developed in the future without necessarily changing how policy specification works.  In addition, embedding security policies in IR (and, furthermore, into executable objects) could serve as a basis for flexible reasoning about more general security policies and run-time deployment of complementary defenses.

\subsection{Policy Instrumentation}%
\label{sub:policy_instrumentation}

The next stage of the \sysname pipeline is responsible for enforcing an IR-embedded policy.  This stage is designed as an LLVM transformation pass that instruments an IR using the embedded policy metadata produced in the previous stage.  This instrumentation pass considers both partition assignments for data as well as privilege management involved in cross-partition control flows.  The nature of the instrumentation also differs based on whether partition assignments and privilege management can be easily resolved at compile time, or whether run-time assistance is required.  We discuss each of these instrumentation classes in the following.  We also note that this pass satisfies our design goal of preventing the accidental introduction of privilege leaks due to partition specification errors, up to the correctness of the pass~\ref{goal:correctness}.

\paragraph{Variable Partitions.}%
\label{par:variable_partitions}
The first and simplest instrumentation class concerns placing variables in the correct partition.  Since all variables must be labeled with policy metadata stating which partition they belong to, instrumentation is a straightforward process of marking each variable as part of the corresponding partition's data section in the final executable object.  Note that this alone does not enforce a partition boundary.  However, grouping data into partitions when laid out in memory during executable loading allows the run-time support library to easily implement that boundary~(\S\ref{sub:run_time_support}).

\paragraph{Dynamic Memory Partitions.}%
\label{par:dynamic_memory_partitions}
In \sysname, dynamically-allocated data must be located in the correct partition.  Since standard memory allocation APIs have no provision for controlling this behavior, the instrumentation pass must dynamically dispatch all dynamic memory operations (e.g., allocation, deallocation) to the correct partition by interposing on those operations at run-time.

The pass first identifies all calls to a dynamic memory API (e.g., \texttt{malloc}, \texttt{operator new}).\footnote{\sysname also supports user-defined allocators labeled with the Clang-supported \texttt{alloc\_size} attribute.}  Pointer argument or return value def-use chains are followed to resolve variable partition assignments---since LLVM IR is in SSA form, this is straightforward to perform.  This associates a partition with each memory operation, and allows the pass to statically dispatch the operation to a partition-aware equivalent that is provided by the run-time support library~(\S\ref{sub:run_time_support}).

The approach currently adopted by \sysname does not support multiple partitions at a single memory operation call site.  This would manifest during pass execution as resolving multiple partition assignments when following def-use chains from an operation.  If this occurs, the stage terminates processing and reports the error.  This is not a fundamental limitation, since more sophisticated instrumentation could propagate a partition label along with the variable at a small run-time cost.  However, we have so far not found this necessary.

\begin{figure}[h]
    \centering
    \includegraphics[width=\linewidth]{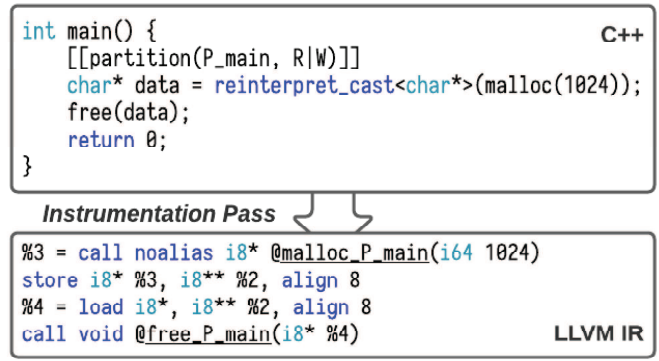}
    \caption{Example of partition assignment and dynamic memory operation dispatch.  Assigning \texttt{data} to \texttt{P\_main} has the effect of redirecting the call to \texttt{malloc} to \texttt{malloc\_P\_main} after instrumentation, which will allocate memory from a partition-specific heap.}
    \label{fig:instrumentation}
\end{figure}

An example of assigning a variable to a partition and dynamic memory operation dispatch is shown in Figure~\ref{fig:instrumentation}.

\paragraph{Static Privilege Modification.}%
\label{par:static_privilege_modification}
Cross-partition function calls require privilege modification from the caller's rights to those of the callee.  Whether this involves replacement or augmentation as specified in the source-level policy, the pass must instrument the function call with the requisite rights management code.  In cases where the complete set of call target(s) can be statically resolved and this set all has identical callee partition rights, the pass can in turn statically inject the necessary privilege modifications.

The pass first injects a preceding call to a run-time support function \texttt{set\_privileges} with the callee's rights at the original call site.  Then, if the original cross-partition call is not marked with the \texttt{noreturn} attribute, a matching \texttt{set\_privileges} call is injected after the original cross-partition call to restore the caller's rights.  Similar instrumentation is performed for exceptional control flows represented in the IR.  However, in cases where the caller and callee's rights are identical, rights management can be statically elided.
\begin{figure}
    \centering
    \includegraphics[width=\linewidth]{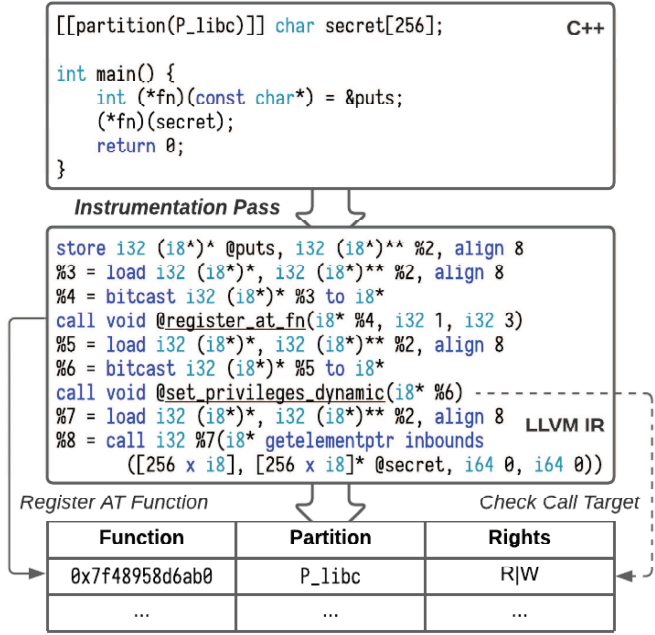}
    \caption{Example of address-taken (AT) function tracking.  \sysname's instrumentation pass registers all AT functions with the run-time.  Before any indirect call is performed, the run-time first checks the AT function table to determine whether privilege modification is necessary.}
    \label{fig:at_tracking}
\end{figure}

\paragraph{Dynamic Privilege Modification.}%
\label{par:dynamic_privilege_modification}
In principle, a function pointer can invoke any target with a compatible function signature, potentially invoking many different targets. As the implementations of these functions might require privileges for different partitions, this poses a challenge for \sysname. Specifically, in these cases, it can be difficult to statically determine what rights to assign for a cross-partition call, or to determine whether an indirect call can cross partitions at all.

While LLVM supports interprocedural alias analyses such as SVF~\cite{sui_2014_detectingmemoryleaks} that can be used to compute precise call graphs, alias analysis is known to be NP-hard and thus can take non-trivial time to compute~\cite{horwitz1997precise}.  Thus, for the initial design we have elected instead to adopt a conservative dynamic solution.  That is, \sysname identifies every location where a function address is assigned to a variable---i.e., the function has its address taken.  Then, the pass instruments the variable assignment with an immediately dominating call to a run-time support library helper function called \texttt{register\_at\_fn}.  This function dynamically tracks all address-taken~(AT) functions and their associated privileges.

The pass then instruments all indirect calls with an immediately dominating call to the run-time support library-provided \texttt{set\_privileges\_dynamic}.  Instead of statically encoding privileges as with \texttt{set\_privileges}, this dynamic analogue consults the dynamically-tracked AT function privilege table to determine the proper callee rights to set.
\subsection{Run-Time Support}%
\label{sub:run_time_support}

\sysname's run-time component has several responsibilities.  First, it initializes policy-specified partitions prior to execution of the main program.  Second, it provides primitives for privilege management for cross-partition control flows.  For both of these tasks, the run-time relies on Intel MPK hardware support.  Finally, it interposes on dynamic memory operations and tracks potential indirect call targets to correctly set privileges for cross-partition indirect calls that could not be resolved at compile-time.

\paragraph{Partition Initialization.}%
\label{par:partition_initialization_}
During \sysname's instrumentation pass, variables are placed in sections corresponding to their partition.  Before the main program executes, the \sysname run-time labels each partition's memory region with the policy-specified label.  This label directly corresponds to a \emph{protection key}, of which 16 are available in the current generation of MPK implementations.

On Linux, this amounts to allocating protection keys with \texttt{pkey\_alloc} and invoking the \texttt{pkey\_mprotect} system call, which extends the familiar \texttt{mprotect} system call with a protection key parameter, on each partition.

\paragraph{Privilege Management.}%
\label{par:privilege_management_}
The \sysname run-time also provides the \texttt{set\_privileges} function that cross-partition call instrumentation invokes to set callee privileges and restore caller privileges.  This function in turn wraps the WRPKRU instruction, which sets the privilege bit vector (R/W bits for each partition) in the PKRU register for the current thread.

\paragraph{Dynamic Memory Operation Dispatch.}%
\label{par:dynamic_memory_operation_dispatch_}
As described previously, the instrumentation pass redirects all dynamic memory operations to partition-aware equivalents provided by the run-time.  For example, \sysname provides \texttt{void* pkey\_malloc(size\_t size, int pkey)} that allocates the requested memory from a heap isolated in the named partition.

While this scheme works for instrumented translation units, dynamic memory operations located in non-instrumented code such as pre-compiled shared libraries require a different approach.  In this case, the run-time relies on link-time interposition---that is, the linker is instructed to redirect calls to functions such as \texttt{malloc} and \texttt{free} to the run-time-provided versions.  For these call sites, the containing object partition label is assumed and attributed using stack inspection rather than expecting the partition label to be supplied as a parameter.  Note that this mechanism---i.e., \verb+ld+'s \verb+--wrap+ option---operates purely at link-time and does not require any run-time intervention, e.g., LD\_PRELOAD.

The \sysname run-time also ensures that memory is scrubbed when it is deallocated by the program.  This is done to avoid unintentional data leaks from partition heaps where, e.g., a buffer holding sensitive data is reallocated and explicitly released to another partition.

\paragraph{AT Function Tracking.}%
\label{par:at_function_tracking_}
The final task \sysname's run-time performs is AT function tracking to correctly support cross-partition indirect calls.  The instrumentation pass allows the run-time to collect all possible indirect call targets---i.e., those that are address-taken---and track these in a table along with their policy-specified privileges.  The instrumentation also allows the run-time to interpose on potential cross-partition indirect calls, performing a lookup into the AT privileges table to assign the correct partitions.  Figure~\ref{fig:at_tracking} steps through a concrete example of this procedure.

\FloatBarrier

\section{Implementation}
\label{sec:impl}

The \sysname prototype is implemented as three main components:
\begin{inparaenum}[\itshape (i)\upshape]
    \item policy embedding modifications to the Clang C++ front-end;
    \item policy enforcement instrumentation as an LLVM transformation pass; and,
    \item the run-time support library.
\end{inparaenum}
Policy embedding was implemented with 154 SLOC (C++ and LLVM TableGen) across 12 files to add support for \sysname pragmas and attributes.  Policy enforcement instrumentation was implemented as an LLVM module-scope transformation pass, and consists of 2,075 SLOC (again, C++ and LLVM TableGen).  It is also responsible for generating a linker script that controls the memory layout of the final program executable.  Clang CFI was used to prevent code reuse attacks against \sysname policies.  Finally, the run-time support library is currently generated via meta-programming by 390 SLOC (C++) in the instrumentation pass.  This approach allows \sysname to tailor each run-time instance to each application.  Dynamic memory management is implemented using jemalloc as a backend, though in principle any compatible allocator such as ptmalloc2 or tcmalloc could be used in its place.

\section{Evaluation}
\label{sec:eval}

\newcommand{\speccpu}[0]{SPEC CPU~2017\xspace}
\newcommand{\speccpurate}[0]{SPECrate~2017 Integer\xspace}

The goal of this evaluation is to measure \sysname in order to answer three broad questions:
\begin{inparaenum}[\itshape (i)\upshape]
    \item \textbf{Security}: Does \sysname harden programs to a similar level as comparable privilege separation or data isolation schemes?
    \item \textbf{Correctness}: Does \sysname preserve the intended behavior of protected applications?
    \item \textbf{Performance}: What is the time and space overhead introduced by \sysname?
\end{inparaenum}

To answer these questions, we constructed a testbed and a set of test programs.  The experimental testbed consisted of a server with an Intel Xeon Gold 6242 CPU @ 2.80~GHz and 64~GB of DDR4 ECC RAM at 2934~MT/s.  This configuration was chosen primarily for its support for Intel Memory Protection Keys~(MPK), which was introduced in certain Skylake processors in 2017.

The first test set is a regression test suite that was developed alongside \sysname.  This test suite contains 23~programs that check the correctness of various aspects of the prototype design and implementation, including memory allocation dispatch, privilege management, and AT function privilege tracking.  Due to space constraints, we do not report further on this test set in this paper.  The other test set we consider in this evaluation is \speccpu, which we discuss in \S\ref{sub:spec_cpu_2017}.  We consider the application of \sysname to real applications in two case studies: Nginx~(\S\ref{sub:nginx}) and SQLite~(\S\ref{sub:sqlite}).  These experiments enable a direct comparison to prior defenses based on Intel MPK~\cite{vahldiek-oberwagner_2019_erimsecureefficient}.  Finally, we measure \sysname's size impact across \speccpu, Nginx, and SQLite in \S\ref{sub:size_overhead}.

\subsection{\speccpu}%
\label{sub:spec_cpu_2017}

\begin{figure}[h!]
    \centering
    \includegraphics[width=0.7\linewidth,angle=270]{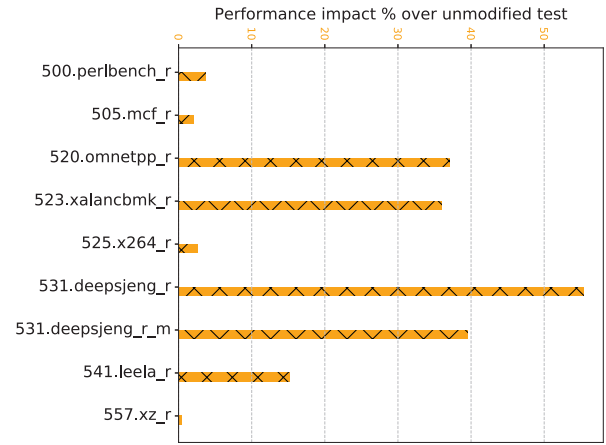}
    \caption{Impact of using a \texttt{module} level protection in Integer \speccpu. Switching occurs automatically to every out of module call, since this a completely unguided policy affecting all computational units indiscriminately the impact can be quite high. Even slight developer guidance selecting modules to be excluded can have a significant performance impact (e.g., \textsf{531\_sjeng\_r.m}).}
    \label{fig:spec}
\end{figure}

\speccpu ships with 43~benchmarks organized into four suites.  In this experiment, we used the 11~benchmarks comprising the \speccpurate suite as a baseline.  We compare this baseline to benchmark programs that were modified to include two partitions: one for the main program, and another for all libraries used by the main program.  We note that this partitioning is intentionally na\"{i}ve, and leads to many cross-partition control flows that are unnecessary from a security perspective.  In this respect, we consider this an approximate worst-case scenario.

While the \speccpurate suite provides 11~benchmarks, we were unable to include two of them in this experiment.  \textsf{548.exchange2\_r} is authored in Fortran and hence cannot be protected by \sysname.  \textsf{502.gcc\_r} could not be compiled using our prototype because our toolchain introduces external symbols in multiple translation units, triggering a known problem with this test that can arise even in unmodified versions in other platforms.\footnote{\url{https://www.spec.org/cpu2017/Docs/benchmarks/502.gcc_r.html}}  All other benchmarks were compiled using the prototype toolchain without problems.

\speccpu includes a validation step that automatically determines whether a benchmark produces the expected output.  This allows us to measure not only the run-time induced by \sysname in this particular partitioning scheme, but also to verify the correctness of the separated program.

Figure~\ref{fig:spec} presents the results of this experiment using the provided test load over 10~iterations to reduce noise.  Overall, it is clear that \sysname introduces a variable impact across the suite.  In four cases, run-time overhead ranges between 1-4\%, while the remainder ranges from 15-55\%.  Investigating this phenomenon revealed that the structure of the program with respect to the na\"{i}ve partitioning scheme selected for this experiment plays a large role in incurred overhead.

Consider \texttt{531.deepsjeng\_r}, a chess-playing application that reports the worst-case overhead of 55\% .  This benchmark makes a large number of calls to library functions (e.g., libc's \texttt{sprintf}), each of which represents a cross-partition control flow and thus requires privilege modification.  This highlights a potential pitfall of na\"{i}ve or automatic partition boundary specifications that do not take a program's call graph and security requirements into account.  With a small manual effort, we identified two functions in the main program partition (\texttt{move} and \texttt{unmove}) that, if placed instead in the library partition, led to a 14\% reduction in run-time overhead.  This illustrates that developer guidance of security policy specification can be an important factor in keeping performance impact to a minimum, so long as desired application-specific security properties are maintained.

Finally, we note that we have not invested significant effort in optimizing our prototype implementation.  There are a number of natural optimization steps that could be taken---e.g., inlining privilege modifications at cross-partition calls, or using an interprocedural alias analysis to reduce the number of indirect calls that must be tracked---that would further reduce \sysname's run-time overhead across the board.
\subsection{Case Study: Nginx}%
\label{sub:nginx}

\begin{figure*}[h]
    \begin{subfigure}[t]{0.49\textwidth}
        \centering
        \includegraphics[width=\linewidth]{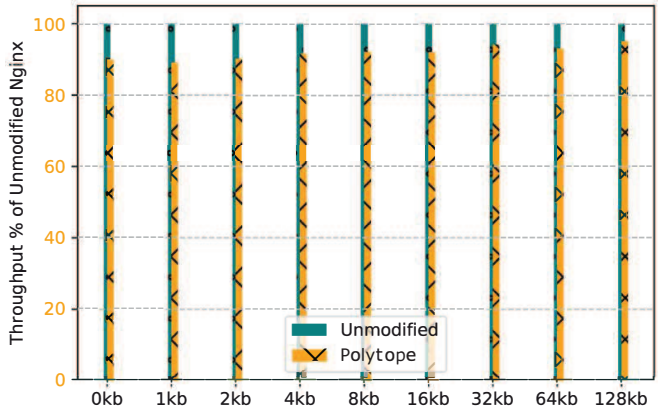}
        \caption{Nginx experiment over localhost (no delay).}
        \label{fig:nginx_nodelay}
    \end{subfigure}
    \hfill
    \begin{subfigure}[t]{0.49\textwidth}
        \centering
        \includegraphics[width=\linewidth]{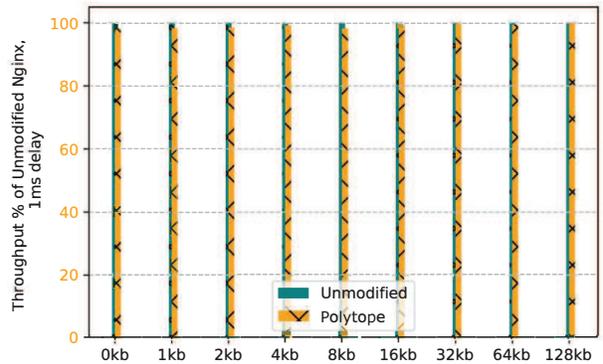}
        \caption{Nginx experiment over localhost (1~ms induced delay).}
        \label{fig:nginx_1msdelay}
        \end{subfigure}
    \caption{Throughput of \sysname-augmented Nginx versus an unmodified baseline version.  The maximum observed impact of our approach (10.1\%) occurs when testing over localhost with no delay and with a 0~kB page load. The absence of network-induced noise or file transfer delays isolates the impact of our approach. In contrast, when testing with a simulated network latency of 1~ms, \sysname's impact is overshadowed by network latency which is orders of magnitude greater than, e.g., privilege management operations (ms vs.~ns).}
\end{figure*}

The objective of this case study is to characterize \sysname's security guarantees and performance impact in the context of prior work when securing a real-world program.  To that end, we selected Nginx (v1.13.1), a popular web server, and compiled it against a version of OpenSSL (v1.1.0.f) known to be vulnerable to Heartbleed (CVE-2014-0160).  We then partitioned Nginx using ERIM's Trusted and Untrusted module approach~\cite{vahldiek-oberwagner_2019_erimsecureefficient}.  In this scheme, memory assigned to libcrypto---the library responsible for providing cryptographic primitives for OpenSSL---is only accessible from Nginx.

In contrast with manual implementation, \sysname policy annotations amounting to 10~SLOC were sufficient to replicate ERIM's protection scheme in terms of partition and privilege assignments.  In contrast, a source-level diff against OpenSSL indicates that ERIM touches 22~files, inserting 538~lines and deleting~744.  The resulting \sysname-protected executable reported 10.1\% latency over baseline in a worst-case experimental configuration that factors out network latency (Figure~\ref{fig:nginx_nodelay}).

\subsection{Case Study: SQLite}%
\label{sub:sqlite}

\begin{figure}
 \centering
    \vspace*{-1.0cm}
    \includegraphics[height=10cm]{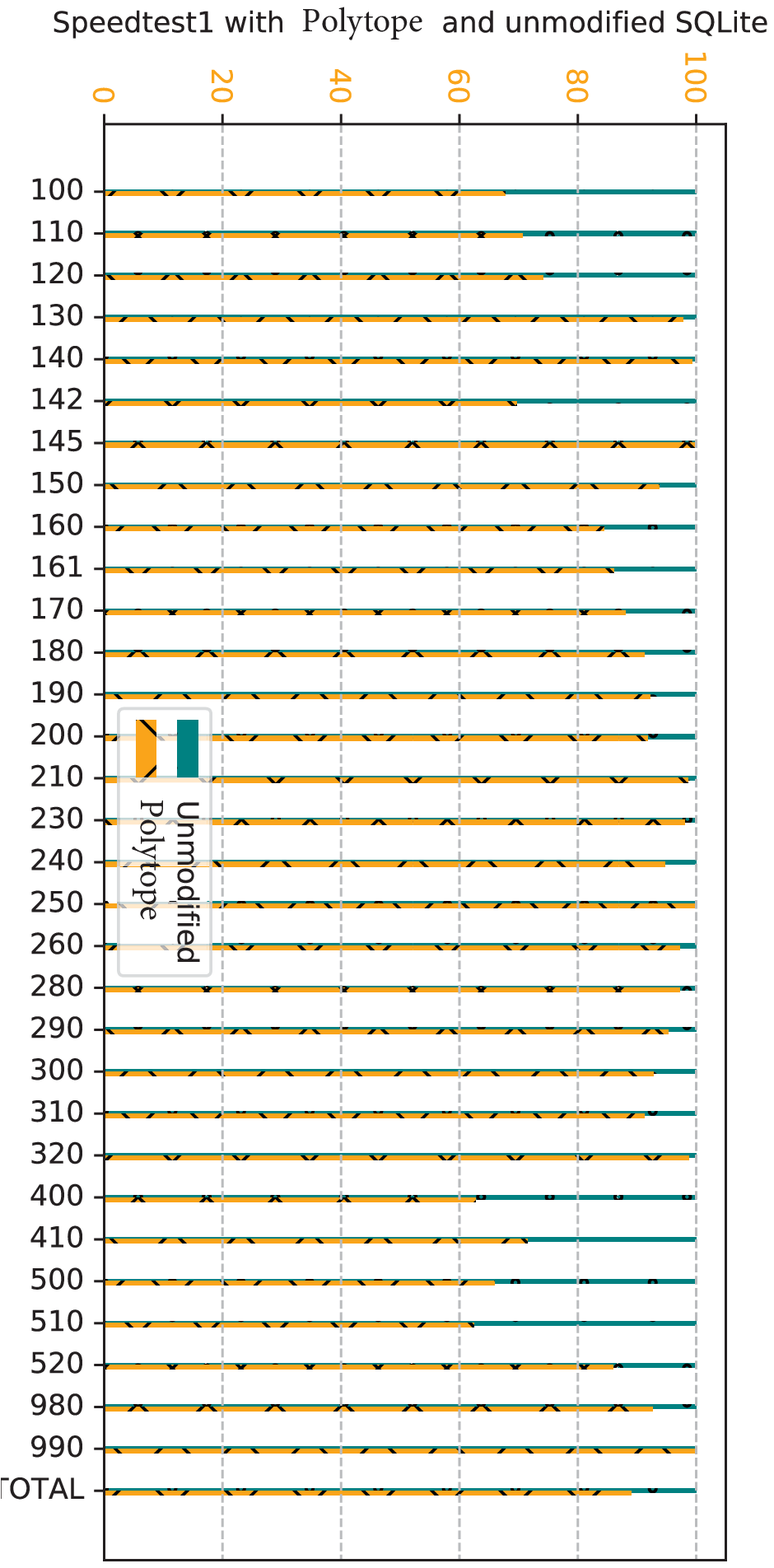}
    \caption{Evaluating performance impact on SQLite using \sysname. We employ the bundled speedtest1 test suite and instruct \sysname to perform a partition switch on every call to libsqlite. The aggregated average impact on all test cases over 20~runs is 11.1\%.}
 \label{fig:sqlite}
\end{figure}

With similar goals as Nginx, we consider SQLite as another real-world case study.  Here, we use speedtest1, a test suite included with SQLite (v3.35.4), as a baseline.  We then create a \sysname-protected version that defines two partitions, one for the test driver and another for libsqlite.

Figure~\ref{fig:sqlite} depicts the results of this experiment.  Mean run-time overhead is measured at 11.1\%, though again there is non-trivial variance across individual benchmarks.  Investigating further, we found that large numbers of cross-partition calls were the main contributing factor to reported overhead.  For instance, tests 400, 410, 500, and 510 report run-time overheads in the 32-38\% range.  Each of these tests issues 70K select, insert, and update queries in rapid succession, which naturally results in a large number of cross-partition calls from the main program into libsqlite.  This again demonstrates that developer guidance is important in defining partition boundaries that balance security and performance.  We note that decoupling policy specification from implementation can help developers more easily test different program partitionings without having to engage in more involved refactoring.
\subsection{Size Overhead}%
\label{sub:size_overhead}

\begin{table*}[!t]
    \begin{tabularx}{\textwidth}{Xrrrrr}
        \toprule
        \textbf{Application} & \textbf{Baseline (kB)} & \textbf{\sysname (kB)} & \textbf{Overhead} & \textbf{Inst.~Only (kB)} & \textbf{Inst.~Contribution} \\
        \midrule
500.perlbench\_r & 2310.60 & 2374.83 & 1.03x & 50.23 & 78.20\% \\
505.mcf\_r & 42.73 & 72.13 & 1.69x & 15.39 & 52.35\% \\
520.omnetpp\_r & 2033.40 & 2070.44 & 1.02x & 23.05 & 62.23\% \\
523.xalancbmk & 4354.01 & 4772.91 & 1.10x & 404.90 & 96.66\% \\
525.x264\_r & 635.47 & 696.44 & 1.10x & 46.97 & 77.04\% \\
531.deepsjeng\_r & 111.33 & 140.94 & 1.27x & 15.61 & 52.72\% \\
541.leela\_r & 170.48 & 209.12 & 1.23x & 24.64 & 63.77\% \\
557.xz\_r & 205.81 & 242.12 & 1.18x & 22.30 & 61.42\% \\
        \midrule
nginx & 1218.14 & 1258.08 & 1.03x & 25.94 & 64.95\% \\
speedtest1 (SQLite) & 58.66 & 75.01 & 1.28x & 2.35 & 14.37\% \\
\midrule
\textbf{Mean}
    & 1114.06
    & 1191.20
    & 1.19x
    & 63.14
    & 62.37\%
\\
\textbf{Median}
    & 420.64
    & 469.28
    & 1.14x
    & 23.84
    & 63.00\%
\\
\textbf{Std.~Dev.}
    & 1414.07
    & 1517.25
    & 0.20x
    & 120.92
    & 21.47\%
\\
        \bottomrule
    \end{tabularx}
    \caption{Size measurements for each program in the test set. The \sysname column reports code total size, including the run-time support library.  Instrumentation size is reported separately, as well as its contribution to total overhead.  We note that relative contributions of instrumentation and run-time can vary significantly depending on program design and, in principle, policy.  For the test set, overhead ranges from 1.02-1.69x, though the mean and median suggest that the overhead skews towards the lower end of this range.}
    \label{table:instrumentation}
\end{table*}

The final measurement we report on in this evaluation is \sysname's size impact on \speccpu, Nginx, and SQLite.  Recall that \sysname's LLVM instrumentation pass adds code to manage privileges for cross-partition calls.  The pass also adds instrumentation to mediate indirect calls and dynamic memory operations by redirecting those to the run-time support library.  The run-time itself also contributes modest overhead that, due to the metaprogramming approach adopted by the prototype, can vary depending on the base program.  Table~\ref{table:instrumentation} depicts the size overhead for each benchmark in \speccpurate as well as Nginx and SQLite.
\subsection{Evaluation Summary}%
\label{sub:evaluation_summary}

\begin{itemize}
    \item \sysname partitions the \speccpurate test suite while preserving benchmark correctness.
    \item \sysname provides equivalent security to prior data isolation work using Intel MPK, with significantly less developer overhead (e.g., 10 SLOC in policy annotations vs.~a diff against OpenSSL consisting of 538~insertions and 744~deletions across 22~files).
    \item Measurements suggest modest run-time overhead that is comparable to, though greater than, prior work, satisfying our performance design goal~\ref{goal:performance}.
    \item Developer input is an important factor in defining program partitionings that represent an optimal point in the security vs.~performance space.  \sysname's design principle of decoupling policy from implementation enables agile exploration of this space without potentially cumbersome refactoring.
    \item \sysname introduces small size overhead for protected programs, though the magnitude of this depends on the number of partitions and cross-partition calls.
\end{itemize}

\section{Discussion}
\label{sec:discussion}

In this section, we anticipate several natural questions regarding this work.

\paragraph{What if developers leak protected data?}%
\label{par:leak_protected_data}
The results of computation involving sensitive data sometimes must be transferred across partition boundaries.  Developers are ultimately responsible for writing code that does not unintentionally violate application security requirements.  Information flow control would allow developers and \sysname to directly reason about dangerous flows.  However, the annotation requirements for an IFC system are substantially higher than that required by this system.  Regardless, an interesting future direction would be to compose \sysname with a lightweight form of IFC.

\paragraph{Why not use memory-safe languages?}%
\label{par:memory_safe_languages}
Memory-safe languages can achieve similar data isolation goals as \sysname, and we agree that they should be used wherever functional requirements permit their use.  However, history informs us that compilers and language run-times have bugs that can result in security vulnerabilities.  So, in the tradition of defense-in-depth, composing language-based security with hardware protection is likely to result in safer systems.

\paragraph{What about sandboxing access to external resources?}%
\label{par:what_about_external_resources}
Privilege separation often encompasses not only data isolation but also restrictions on access to resources external to a protection domain.  For instance, on Linux this might correspond to changing process ownership to a non-root user, only granting specific capabilities to a process, or enforcing SECCOMP-BPF filters on system calls issued by a process.  The common denominator among these examples is that they operate on a process basis, whereas \sysname security boundaries are based purely on memory regions.  Devising new sandboxing primitives appropriate to \sysname partitions is an avenue of future work we intend to explore.

\paragraph{Annotations are onerous.}%
\label{par:annotations_are_onerous}
We agree that they can be!  Thus, in designing \sysname we emphasized minimizing the number of annotations required to cover common use cases, and took care to integrate policy specification with the attribute language feature introduced in the C++11 standard.

\section{Related Work}
\label{sec:related}

\sysname builds on an extensive literature on privilege separation to which we compare and contrast in the following.

\paragraph{Privilege separation.}%
\label{par:privilege_separation}

Process-based separation~\cite{kilpatrick_2003_privmanlibrarypartitioning,brumley_2004_privtransautomaticallypartitioning, liu_2019_programmanderingquantitativeprivilege,barth_2008_securityarchitecturechromium,mozillafoundation_2021_electrolysis} achieves data isolation by separating application functionality into multiple processes.  The OS enforces hardware-backed memory isolation between processes, serving as strong security boundary largely preventing unprivileged processes from affecting each other or the system.  These systems commonly require invasive refactoring of the program into a multi-process architecture separating privileged and unprivileged processes.  This partitioning is either performed manually~\cite{kilpatrick_2003_privmanlibrarypartitioning,brumley_2004_privtransautomaticallypartitioning}, semi-automatically aided by tools~\cite{gudka_2015_cleanapplicationcompartmentalization}, or fully automatically using a performance budget~\cite{liu_2017_ptrsplitsupportinggeneral,liu_2019_programmanderingquantitativeprivilege}.  IPC traditionally incurs sizable run-time overhead.  In contrast, intra-process privilege separation based on new hardware primitives such as Intel MPK significantly reduce the overhead associated with privilege separation.

Prior work in this space has made use of language, compiler, or run-time loader extensions.  SOAAP~\cite{gudka_2015_cleanapplicationcompartmentalization}, for instance, uses annotations to define \emph{compartmentalization hypotheses} for process-separated programs that can be accepted or rejected.  These hypotheses can express whether code respects a hypothesized isolation policy, or whether a policy can satisfy a declared performance budget.  These hypotheses are checked by an LLVM-based analysis.  However, SOAAP hypotheses are oriented towards reasoning about policies rather than enforcement.  They also only consider process-based isolation schemes.

Elfbac~\cite{bangert_2013_elfbacusingloadera} defines a policy language that can express a coarse-grained intra-process isolation policy embedded in ELF executables.  While similar in spirit to \sysname, to our knowledge Elfbac has not undergone a performance evaluation.  However, its reliance on software-based shadow memory contexts instead of a hardware mechanism like MPK makes it likely that it is unlikely to exhibit competitive performance.  It also does not adopt the source-level security policy approach adopted by \sysname.

Ghosn et al.~define \emph{enclosures}~\cite{ghosn_2021_enclosureslanguagebasedrestrictions}, which equip language closures with a memory view and system call policy.  Enclosures have been implemented for Python and Go, and are enforced using an accompanying sandbox based on either Intel VT-x or Intel MPK.  To our knowledge, enclosures have not been implemented for a lower-level language like C++, nor are we aware of a published performance evaluation of the approach.

RLBox~\cite{narayan_2020_retrofittingfinegrain} builds on the C++ type system to enforce either SFI (e.g., NaCL, WebAssembly) or process-based isolation.  Notably, it has been used to isolate libgraphite using a WebAssembly sandbox in production Firefox.  RLBox can also enforce system call policy on a process granularity using SECCOMP-BPF.  However, it is limited to library isolation, whereas \sysname is capable of much finer-grain isolation policies.

\paragraph{Intel MPK.}%
\label{par:intel_mpk}

Systems relying on Intel MPK to achieve data isolation are most closely related to our work.  ERIM~\cite{vahldiek-oberwagner_2019_erimsecureefficient} achieves low cost data isolation---e.g., less than 5\% to protect private keys for Nginx---by separating a binary into trusted and untrusted compartments.  Developers can use ERIM to achieve fine-grained control over trusted partition data by manually switching access rights through specialized call gates.  Libmpk~\cite{park_2019_libmpksoftwareabstraction} abstracts MPK, virtualizing the limited number of protection keys and preventing key use-after-free vulnerabilities.  Donky~\cite{schrammel_2020_donkydomainkeys} extends the RISC-V instruction set architecture with functionality substantively similar to MPK, but supports up to 1,024 keys.  A common denominator for all aforementioned systems is their reliance on their respective specialized software APIs that entail extensive and error-prone developer effort to implement policies.  \sysname, on the other hand, provides a concise policy language that can express equivalent data isolation guarantees while avoiding the expression of privilege leaks.

Recent work by Connor et al.~\cite{connor_2020_pkupitfallsattacks} has shown that MPK-based isolation can be bypassed in some cases due to missing permission checks in the Linux kernel.  A secure \sysname policy depends on proper enforcement of protection keys in the underlying OS kernel that, while orthogonal to this work in principle, is nevertheless important.  We anticipate that such loopholes in kernel memory protection policy enforcement will be closed as they are found, but automatically discovering other MPK bypasses is an interesting line of future work.

\paragraph{OS-based systems.}

As thread, memory, and resource management is handled by the OS, a natural foundation to implement support for in-process privilege separation is the kernel.  Systems in this area require kernel modifications or kernel modules ~\cite{bittau_2008_wedgesplittingapplications,efstathopoulos_2005_labelseventprocesses,witchel_2005_mondrixmemoryisolation,hsu_2016_enforcingleastprivilege,mambretti_2016_trellisprivilegeseparation,belay2012dune,hedayati_2019_hodorintraprocessisolation,koning2017no,litton2016light} to implement their systems.  They offer significantly better runtime performance than process-separated approaches and the omission of kernel mediation such as in Dune~\cite{belay2012dune} can further improve performance.  \sysname, in contrast, does not rely on any kernel or system wide modifications, and merely requires a handful of annotations of variables and code.  However, since \sysname relies and builds on top of existing system components it induces higher performance impact than what can be achieved through system and kernel modification.

\paragraph{SFI.}
\label{par:sfi}

Software fault isolation~\cite{wahbe1993efficient,ford2008vx32,yee2009native,deng2015isboxing} can also isolate memory regions according to developer-provided constraints.  SFI systems commonly instrument binaries coupled with CFI techniques to limit memory accesses access to a set of allowed targets.  A recently proposed system called DynPTA~\cite{palitdynpta} uses source code annotations to establish protected compartments and encrypts protected memory using AES.  DynPTA ensures that only authorized compartments can access protected memory.  As memory is only decrypted when required, DynPTA provides strong guarantees even in the presence of Spectre attacks.  However, DynPTA's performance overhead is correlated with the volume of protected data and the frequency with which this data is accessed.  This overhead thus varies significantly, between 19.2\% for Nginx to 400\% for a fully protected implementation of mergesort.  In contrast, \sysname and MPK-based approaches do not perform expensive encryption operations and only require fast register operations to switch permissions irrespective of the volume of protected data.

\paragraph{Hardware enclaves.}
\label{par:sgx}

Data isolation can also be achieved using recent hardware additions such as Intel SGX~\cite{costan_2016_intelsgxexplained} and Arm TrustZone~\cite{arm_2020_trustzone}. In such schemes sensitive data and computation takes place in protected enclaves, communicating through protected endpoints with the rest of the system. Such approaches however face communication latency, limited memory size, and the inability to dynamically allocate memory~\cite{costan_2016_intelsgxexplained,koning2017no}.  In contrast, \sysname does not have memory limitations, supports dynamic memory operations, and cross-partition communication simply requires an update to the PKRU register which costs a few CPU cycles.

\section{Conclusion and Future Work}
\label{sec:conclusion}

In this paper, we presented \sysname, a language extension to C++ that allows for concise expression and enforcement of privilege separation policies for data isolation.  \sysname takes the form of a modified Clang front-end to embed policies in LLVM IR as metadata nodes.  An LLVM pass interprets an embedded policy by placing variables in Intel MPK-secured memory regions.  Cross-partition calls and dynamic memory allocations are instrumented to enforce declared security policy.  Control-flow integrity ensures that declared policies cannot be subverted via code reuse attacks such as return- or jump-oriented programming.  A run-time support library manages partitions and protection keys.  Our evaluation demonstrates that \sysname can enforce equivalent policies to prior work with a low annotation burden and comparable performance overhead while preventing the expression of privilege leaks.

We envision \sysname as a starting point for making privilege separation and secure software design more accessible to developers.  In future work, we plan to build on \sysname in several complementary directions.  These include improving the expressiveness of \sysname policies, e.g., adding the ability to restrict access to external resources on a partition basis; adding additional guarantees on data flows across partitions; improving support for statically checking properties of compositions of policies; and, integrating support for emerging hardware security primitives such as isolated memories.

\printbibliography
\appendix

\section{Example Abstract Policy}%
\label{sec:example_abstract_policy}

Recall the example application shown in Listing~\ref{list:example}.  Assuming that the modules comprising the program are the main program itself, libc, and OpenSSL, a partial abstract \sysname policy is as follows.\footnote{This example is simplified for exposition.  The full program would include more libraries and, in turn, partitions.  In addition, \(\alpha\) would be defined on variables contained in modules aside from \textsf{main}, while \(\pi\) would be defined on statements outside of \textsf{main} as well.}

\begin{align*}
    P &= \big\{ \; \partmain,\partlibc,\partlibssl \; \big\} \\
    \phi(p) &= \left\{
        \begin{array}{c}
            \left(\partmain,\left\{\privread,\privwrite\right\}\right), \\
            \left(\partlibc,\left\{\privread,\privwrite\right\}\right), \\
            \left(\partlibssl,\left\{\privread,\privwrite\right\}\right)
        \end{array}
        \right\} \\
    \alpha(v) &= \left\{
        \begin{array}{c}
            \left(\mathtt{private\_key},\partlibssl\right), \\
            \left(\mathtt{state},\partmain\right), \\
            \left(\mathtt{request},\partmain\right), \\
            \left(\mathtt{response},\partmain\right)
        \end{array}
        \right\} \\
    \pi(s, p) &= \;
        \begin{dcases}
            \; \left\{\privread,\privwrite\right\} & \text{ if } s \neq 6, p = \partmain, \\
            \; \left\{\privread,\privwrite\right\} & \text{ if } s = 6, p = \partlibssl, \\
            \; \emptyset & \text{ otherwise.}
        \end{dcases}
\end{align*}

\end{document}